**AI Tutors vs. Tenacious Myths: Evidence from Personalised Dialogue**

**Interventions in Education**

Brooklyn J. Corbett[1] & Jason M. Tangen[1]

[1]School of Psychology, The University of Queensland, St Lucia 4072, QLD,

Australia.

**Corresponding author:** Brooklyn J. Corbett

**Corresponding author email:** b.corbett@uq.edu.au

**Last revised:** 09/06/25

# Abstract

Misconceptions in psychology and education persist despite clear contradictory evidence, resisting traditional correction methods. This study investigated whether personalised AI dialogue could effectively correct these stubborn beliefs. In a preregistered experiment (N = 375), participants holding strong psychology misconceptions engaged in one of three interventions: (1) personalised AI dialogue targeting their specific misconception, (2) generic textbook-style refutation, or (3) neutral AI dialogue (control). Results showed that personalised AI dialogue produced significantly larger immediate belief reductions compared to both textbook reading and neutral dialogue. This advantage persisted at 10-day follow-up but diminished by 2 months, where AI dialogue and textbook conditions converged while both remained superior to control. Both AI conditions generated significantly higher engagement and confidence than textbook reading, demonstrating the motivational benefits of conversational interaction. These findings demonstrate that AI dialogue can accelerate initial belief correction through personalised, interactive engagement that disrupts the cognitive processes maintaining misconceptions. However, the convergence of effects over time suggests brief interventions require reinforcement for lasting change. Future applications should integrate AI tutoring into structured educational programs with spaced reinforcement to sustain the initial advantages of personalised dialogue.

# Introduction

Misconceptions and myths in psychology and education are both common and stubborn. Myths such as "we only use 10% of our brain" or that teaching is most effective when tailored to "learning styles" remain widespread despite clear evidence to the contrary (Howard-Jones, 2014; Newton & Miah, 2017; Pashler et al., 2009). Even well-designed educational interventions like lectures, handouts, and fact-sheets often fail to uproot deeply felt convictions (Zengilowski et al., 2021), especially when myths align with intuition or identity (Nyhan & Reifler, 2010). This resistance suggests a fundamental challenge for educational practice.

These stubborn misconceptions extend well beyond public misunderstanding. Psychology students and practising professionals frequently endorse erroneous beliefs about their own field. For instance, over 90% of teachers and more than 80% of psychology undergraduates endorse "learning styles" despite extensive research showing no benefits (Dekker et al., 2012; Morehead et al., 2015; Pashler et al., 2009). Likewise, many practising psychologists still administer scientifically unsupported assessment tools, including projective measures such as the Rorschach and human figure-drawing tests (Benson et al., 2019; Baggi & Martino, 2024; Neal et al., 2019).

When professionals who should be guided by psychological science cling to these misconceptions, the consequences ripple outward—shaping therapeutic practice (Dekker et al., 2012), diverting educational resources (e.g., schools and universities have been estimated to spend millions each year on learning-styles materials; Cuevas, 2015), and eroding public trust in the discipline. The persistence of these myths is particularly concerning because they often masquerade as scientific knowledge while lacking empirical support, making them resistant to correction efforts. Why do such myths prove so tenacious?

**Why do Myths and Misconceptions Stick?**

Dual process theory explains much of the resilience of misconceptions. Our thinking can be categorised by two modes: fast, automatic, intuitive responses that arrive effortlessly, and slower, deliberative reasoning that requires conscious effort (Evans & Stanovich, 2013; Kahneman, 2011). When confronted with claims or questions, our immediate intuitions quickly generate answers that feel right—and these intuitive responses are often reinforced by the very features that make myths compelling. Many misconceptions have a veneer of plausibility precisely because they align with our automatic thinking patterns (Thompson et al., 2011). Personal stories and vivid examples often trump abstract statistics in human judgment (Borgida & Nisbett, 1977; Hornikx, 2018). Consider the individual who clings to a false belief about an ineffective medical treatment because they know someone who *seemed* to improve after using it, even if large-sample evidence shows no effect.

Confirmation bias further compounds this challenge—our tendency to selectively seek, notice, and remember information that confirms existing beliefs while dismissing contradictory evidence (Nickerson, 1998). Once established, misconceptions resist correction as people interpret new information to support pre-existing views (Lewandowsky et al., 2012). In short, our cognitive architecture—favouring quick thinking, confirming what we already suspect, and valuing vivid personal evidence—presents barriers to correcting misconceptions.

Not everyone is equally susceptible to these cognitive traps, however (Stanovich & West, 2008; Pennycook & Rand, 2019). Those who habitually engage in analytical thinking are better at discerning false from true information and resisting misinformation, regardless of their initial beliefs (Bago et al., 2020; Pennycook & Rand, 2019; Sultan et al., 2024). This suggests that effective correction requires prompting learners to engage deliberative thinking and reasoning (Schwarz et al., 2016): it should slow people down and engage their analytical faculties to override the intuitive but incorrect responses (Bago et al., 2020; Lewandowsky et al., 2020; Ecker et al., 2022).

**Traditional Refutation Texts—and Their Limits**

Refutation texts are widely used to correct misconceptions across domains from psychology and science education to health communication and economics (e.g., Danielson et al., 2024; Guzzetti et al., 1993; Tippett, 2010; Mason et al., 2017). These texts identify a misconception, state its inaccuracy, and explain the correct concept with evidence (Tippett, 2010). For example, a refutation text about the "10% brain use" myth would explicitly label this belief as false before presenting neurological evidence of whole-brain activity. Decades of research show well-designed refutation materials often outperform standard expository texts in reducing erroneous beliefs (See Chan et al., 2017; Danielson et al., 2024; Walter & Murphy, 2018, for reviews). Even brief refutations can correct beliefs in the short term (Ecker et al., 2019; Mason et al., 2017).

Despite these successes, refutation texts show inconsistent effectiveness across individuals and contexts. They present the same argument to all readers regardless of individual reasons for holding beliefs or prior knowledge (Danielson et al., 2024; Walter & Murphy, 2018; Zengilowski et al., 2021). Consequently, generic debunking messages often fail to achieve lasting change (Zengilowski et al., 2021), particularly for those strongly committed to misconceptions (Ecker et al., 2010; Nyhan & Reifler, 2010). Static texts cannot probe deeper when initial arguments fail to convince, adapt explanations to individual understanding, or persist when resistance appears. These limitations suggest that more dynamic approaches to misconception correction may be needed.

**Active Learning and Personalisation**

What elements might create more effective approaches to misconception correction? Research on learning and persuasion points to two key factors: active engagement and personalised feedback. Active engagement significantly enhances knowledge acquisition and revision: people learn more effectively by actively participating—answering questions, explaining concepts, or applying knowledge—rather than passively receiving information (Chi, 2009; Deslauriers et al., 2019; Freeman et al., 2014; Maceiras et al., 2025; Theobald et al., 2020). When applied to misconception correction, this active engagement becomes particularly powerful: having learners confront discrepancies between their beliefs and evidence represents a crucial step in conceptual change (Posner et al., 1982). For example, predict-observe-explain sequences, where someone predicts an outcome and then observes conflicting results, can create productive cognitive conflict that motivates belief revision (Mazur, 1997; White & Gunstone, 1992).

The second critical element is personalisation and feedback—tailoring interactions to individual needs and providing targeted responses (Hattie & Timperley, 2007; Shute, 2008; Van der Kleij et al., 2015) Research consistently shows that personalised one-on-one tutoring produces superior learning outcomes compared to conventional instruction, with students showing greater knowledge gains and deeper understanding (Bloom, 1984; Chi et al., 2001; Nickow et al., 2020; VanLehn, 2011). This approach succeeds by identifying student misconceptions in real-time, adjusting explanations to the individual, and providing immediate feedback (Chi et al., 2001; Kulik & Fletcher, 2016; VanLehn, 2011).

This personalised approach proves particularly valuable for education and misconception correction because individuals hold false beliefs for diverse reasons. One person might believe the "10% brain use" myth due to motivational anecdotes, while another believes it from misinterpreting brain imaging statistics. Effective refutation for the first might appeal to common sense and everyday evidence, whereas for the second, it might explain how neuroimaging works. Recent evidence supports this approach: in a controlled study with educators, personalised refutation texts that addressed each teacher's prior responses led to significantly greater conceptual change than generic refutation or standard texts (Dersch et al., 2022). However, this approach still required pre-designed responses and couldn't provide ongoing adaptive feedback during the interaction.

While such personalised instruction at scale was historically impractical, computer-based intelligent tutoring systems demonstrate the value of adaptivity; these pre-programmed systems that customise problems to student performance show improved learning gains compared to static problem sets (VanLehn, 2011). Yet even these systems remain limited by their predetermined algorithms and cannot engage in natural dialogue or respond flexibly to unexpected misconceptions.

**AI as a Solution**

Recent advances in AI, particularly large language models (LLMs) like ChatGPT, Claude, and Gemini, offer a promising solution for scalable personalised misconception correction—enabling one-on-one dialogues that can challenge false beliefs and provide tailored counterarguments at unprecedented scale. Modern LLMs understand natural language and generate human-like responses, enabling truly interactive dialogues that surpass the capabilities of both static refutation texts and pre-programmed tutoring systems.

The conversational nature of LLMs—their ability to engage in natural, back-and-forth dialogue—makes them particularly promising for belief change. Costello et al. (2024) demonstrated this potential by engaging over two thousand participants holding various conspiratorial beliefs in personalised conversations with an advanced LLM. Their 3-round intervention reduced conspiracy belief by roughly 20% on average, with effects persisting two months later. The personalised AI conversations not only debunked specific conspiracies but also reduced participants' credulity towards other conspiracy theories, suggesting broad corrective impact.

Costello's findings align with broader evidence that LLMs possess significant persuasive capabilities across diverse contexts. In structured debates, LLM-generated arguments were as convincing as those of human experts (Palmer & Spirling, 2023), while role-playing LLMs shifted attitudes on polarised U.S. policy issues as effectively as human experts (Hackenburg et al., 2023). Additional studies support these findings: Breum et al. (2023) found that conversing with an LLM changed political opinions comparable to human persuaders, while Goel et al.

(2024) showed that 29% of participants fully retracted false beliefs after AI dialogue, with the AI proving as persuasive as human peers.

These studies provide compelling evidence for why AI dialogue might effectively address psychology misconceptions. AI dialogue creates personalised cognitive conflict by surfacing inconsistencies in learners' reasoning. When someone contradicts themselves, the AI can point this out immediately ("You said X, but consider that Y is observed instead"), providing targeted feedback precisely when misconceptions surface. The interactive format disrupts the fast, intuitive thinking underlying many psychology myths by encouraging analytical processing and ensuring engagement with counterevidence that might otherwise be dismissed (Evans & Stanovich, 2013; Lewandowsky et al., 2012).

Recent experimental work by Costello et al. (2025) supports this reasoning-based account. In a study that systematically varied the features of AI interactions with conspiracy believers, they identified factual counterevidence as the critical ingredient. When AI was instructed to persuade without providing rational arguments or evidence, the debunking effect was eliminated entirely. Conversely, removing persuasive intent while having AI simply provide factual information maintained effectiveness. These findings suggest that AI dialogue succeeds by engaging classical reasoning processes—people update their beliefs when presented with compelling evidence, even when that evidence challenges deeply held convictions.

However, most research has focused on conspiracy theories, political attitudes, or other ideologically-laden beliefs. Psychology misconceptions may operate quite differently: while conspiracy theories often form interconnected belief systems tied to identity and worldview, psychology myths tend to exist as isolated factual errors picked up through education or popular culture. This structural difference matters because conspiracy theories' interconnected nature may make them more vulnerable to broad challenges (as Costello found with spillover effects), whereas psychology misconceptions may require more targeted, content-specific correction. Whether AI tutoring can effectively address these discrete factual errors—and whether such correction produces the same spillover effects—remains an open question that our study investigates.

**Current Study**

Building on Costello et al.'s (2024; 2025) work with conspiracy theories, we investigate whether the same conversational AI approach can debunk everyday psychology myths. We designed an AI-driven dialogue intervention where participants engage in a one-on-one conversation with an AI tutor that is programmed to refute a target misconception. The AI presents evidence and tailored explanations to the participant, while encouraging them to reflect and ask questions.

We compare this Misconception AI Dialogue against two conditions: a Textbook Refutation condition (participants read static passages debunking the same misconceptions) and a Neutral AI Dialogue control (participants chat with AI about unrelated topics, providing interactivity without corrective content). This design addresses three key research questions: (1) Does a personalised AI debunking dialogue lead to a greater reduction in belief in the misconception than a traditional refutation text or an equivalent interactive session with no corrective content? (2) Do belief changes persist over time? (3) Does AI dialogue foster broader scepticism towards other undiscussed misconceptions?

Ultimately, our study aims not only to test the efficacy of AI tutoring for myth debunking but also to illuminate the mechanisms behind any observed effects. We directly contrast interactive AI dialogue with both static text and neutral AI interaction to isolate the specific value of corrective dialogue. Our goal is to inform both the theory (how dialogue-based persuasion influences cognitive processes) and practice of misconception remediation: if successful, AI tutors could become powerful allies in promoting scientific literacy, enabling learners to unlearn myths and align their intuitive beliefs with evidence-based knowledge in psychology and beyond.

## Materials and Methods

We used the Qualtrics platform for survey administration, adapting a template provided by Costello et al. (2024), and recruited participants through Prolific, an online recruitment platform. Before participating, all individuals read an information sheet about the study and provided informed consent. We obtained ethics approval from the Human Research Ethics Committee at The University of Queensland (Protocol Number: 2024/HE001249). We preregistered the study on the Open Science Framework (https://osf.io/jgksa/?view_only=ab4a541dc6b047a68d76135a87061260) and made all study components, including power simulations, analysis scripts, de-identified data, and methods and materials, available (https://osf.io/wseq3/?view_only=bba17d9c74ca4dfca02a716cb2ed21f6.

**Participants**

We conducted a pilot study with 20 participants per condition (60 total) to inform our power analysis. Using data from our pilot study and observed effect sizes from Costello et al. (2024), we conducted power simulations for various sample sizes. These simulations were run 1,000 times to determine our ability to detect significant differences between the Misconception AI Dialogue, Neutral AI Dialogue, and Textbook Reading conditions. We found that 100 participants per condition for the immediate post-test and 75 per condition for delayed post-tests (accounting for an expected 25% attrition rate) provided optimal statistical power. However, to guard against the chance of higher attrition rates and ensure sufficient statistical power even under less favourable conditions, we recruited 125 participants per condition, allowing us to maintain sufficient power for our follow-up analyses.

Participants were adults aged 18 or older who were fluent in English and had a 98% approval rate on previous Prolific studies. To be eligible for the study, participants had to score above 50% on at least one misconception in the pre-intervention survey, ensuring they held a meaningful belief in misconceptions. We excluded six participants who did not meet this requirement. In total, we recruited 375 eligible participants across the three conditions: Misconception AI Dialogue, Neutral AI Dialogue, and Textbook Reading.

Participants ranged in age from 19 to 78 years ($M = 40.25$, $SD = 13.21$). The sample was 56.8% female (n = 213), 41.6% male (n = 156), and 1.6% non-binary/third gender (n = 6). Nearly half of participants (47.2%) held a bachelor's degree, with the remainder distributed across various education levels (see Table 1).

**Table 1**

*Participant Education Levels*

| Education Level | *n* | % |
| --- | --- | --- |
| Bachelor's degree | 177 | 47.20 |
| Some college, no degree | 83 | 22.13 |
| Master's degree | 52 | 13.87 |
| High school graduate or equivalent | 36 | 9.60 |
| Associate's degree | 11 | 2.93 |
| Doctorate degree (e.g., PhD, EdD) | 11 | 2.93 |
| Professional degree (e.g., MD, JD) | 4 | 1.07 |
| Less than high school | 1 | 0.27 |

## Pre-Intervention Measures

Participants began by completing an initial survey assessing their beliefs in various misconceptions related to cognitive psychology (see Table A.1). This T6-item survey included statements such as, "We only use 10% of our brain's full potential" and "Liars can be easily detected through their body language and facial expression." Participants rated each statement on a 0-100 scale, where 0 represented "definitely false" and 100 represented "definitely true." We developed the survey using large language models (GPT-4o, Claude 3 Opus, Gemini Advanced) and established misconception surveys (Bernstein et al., 2023; Lilienfeld et al., 2011). The survey demonstrated good internal consistency during pilot testing (Cronbach's $\alpha$ = 0.868 at Time 1, 0.833 at Time 2) and reasonable test-retest reliability over a 10-day interval (average item correlation coefficient of 0.733).

After completing the survey, participants were asked to elaborate on their strongest held belief.

*"Among the statements presented, you indicated the strongest belief in the following statement, with a rating of [RATING/100].*

*[STATEMENT].*

*Please explain in your own words why you believe this to be true. For example, are there certain pieces of evidence, events, information sources, or personal experiences that have significantly shaped your view?"*

**Interventions**

After completing the pre-treatment measures, participants were randomly assigned to one of three interventions: one experimental condition (Misconception AI Dialogue) and two control conditions (Neutral AI Dialogue and Textbook Reading). For the two AI dialogues conditions, we used Costello et al.'s (2024) Qualtrics template, which incorporated JavaScript to facilitate real-time AI interaction by calling Claude 3.5 Sonnet's API, dynamically injecting participant-specific information into the model's instructions, and displaying the AI's responses. We adapted their approach by adjusting the system prompt to suit our needs.

In the Misconception AI Dialogue condition, participants engaged in a three-round dialogue with Claude 3.5 Sonnet, where the AI addressed their highest-rated misconception. The AI provided counterarguments, explanations, and evidence to challenge the misconception, encouraging participants to reflect and interact. The dialogue began with a system message that included context, the goal of addressing misconceptions, and specific participant information (e.g., belief strength and reasons for holding the misconception). The previous AI and participant messages were included in rounds 2 and 3 to maintain continuity. The full system prompt for this condition can be found in Table B.1.

In the Neutral AI Dialogue condition, participants were randomly assigned to one of three neutral topics that replicated those used by Costello et al. (2024) and were unrelated to misconceptions. These topics included discussing their experiences with their healthcare system, debating whether they prefer dogs or cats, or discussing their past experiences with firefighters. This condition was designed to maintain participant engagement without providing corrective feedback on misconceptions. The AI interaction followed the same three-round format as the experimental condition to ensure consistency in engagement. We adjusted the prompts for this condition to generate a comparable amount of text as in the Misconception AI Dialogue condition, ensuring uniformity across conditions. The full prompts for this condition can be found in Table B.1.

In the Textbook Reading condition, participants read a factual passage addressing their highest-rated misconception. These passages mimicked traditional educational materials, providing non-interactive information that participants could absorb at their own pace. We designed the content in the style of a cognitive science textbook, indirectly addressing misconceptions using detailed explanations and accurate scientific overviews. We matched the length of these passages to the AI dialogues based on pilot testing to ensure similar engagement duration across conditions, allowing for comparable analyses. The prompt used to generate these passages, as well as an example passage, can be found in Table B.1.

**Post-Intervention Measures**

After completing their assigned intervention, participants completed the same 16-item misconceptions survey used in the pre-intervention phase to assess any changes in their beliefs. Following this, participants answered a series of post-treatment questions, which included items related to their engagement with the intervention and confidence in understanding the discussed topics, each rated on a

0-100 scale. Participants were also asked about their familiarity with generative AI, their level of trust in generative AI, and how frequently they use generative AI tools, each rated on a 7-point Likert scale. Additionally, they provided demographic information including age, gender, and education.

**10-Day and 2-Month Follow-up**

We recontacted participants twice after completing the intervention. The first follow-up occurred 10 days post-intervention (n = 359, dropout rate = 5.6% for the Misconception AI Dialogue condition, 6.4% for the Neutral AI Dialogue condition, and 0.8% for the Textbook Reading condition). Participants who completed the 10-day follow-up did not significantly differ from those who did not return in terms of their pre-intervention belief strength, $t(17.36) = 0.08$, $p = .93$. During this follow-up, participants completed the same misconceptions survey used in the earlier phases of the study. The second follow-up took place 2 months post-intervention (n = 326, dropout rate = 11.2% for Misconception AI Dialogue, 20.0% for Neutral AI Dialogue, and 8.0% for Textbook Reading). As with the 10-day follow-up, participants who completed the 2-month follow-up did not significantly differ from those who dropped out in terms of pre-intervention belief strength, $t(66.62) = -0.56$, $p = .57$. The follow-up survey was identical to that of the 10-day follow-up, focusing solely on the misconceptions survey to measure the durability of belief change.

**Hypotheses**

We tested several hypotheses regarding the effectiveness of different interventions on misconception correction. The primary hypotheses were as follows: (1) Immediate Effect: Participants in the AI Tutoring Condition would show a greater reduction in belief in misconceptions immediately after the intervention compared to those in the Neutral AI and Textbook Reading conditions, with the AI Tutoring Condition expected to be the most effective, followed by the Textbook Reading Condition, and the Neutral AI Condition showing the least change. (2) Long-term Effects: The AI Tutoring Condition would show sustained belief reduction at 10-day and 2-month follow-ups, while the Textbook Reading Condition would show some lasting effect, but less than the AI Tutoring Condition. (3) Generalised Impact: The AI Tutoring Condition would lead to greater reductions in non-targeted misconceptions both immediately and at follow-up, compared to the Textbook Reading and Neutral AI Conditions.

We also hypothesised that both AI conditions (AI Tutoring and Neutral AI) would result in higher participant engagement compared to the Textbook Reading Condition, with the AI Tutoring Condition expected to elicit the highest engagement. Additionally, we hypothesised that participants in the AI Tutoring Condition would report higher confidence in understanding the discussed topics compared to those in the Neutral AI and Textbook Reading conditions.

## Results

### Immediate Reduction in Misconception Beliefs

Can engaging with an AI reduce belief in misconceptions about cognitive psychology? To answer this question, we conducted an Analysis of Covariance (ANCOVA) to examine the immediate effects of our interventions on participants' belief in belief in their highest-rated misconception. We aimed to determine whether the Misconception AI Dialogue would lead to a greater reduction in post-intervention belief ratings compared to the Neutral AI Dialogue and Textbook Reading conditions, while controlling for pre-intervention belief ratings.

With post-intervention belief as the dependent variable, Intervention Type (Misconception AI, Neutral AI, Textbook Reading) as the independent variable, and pre-intervention belief as the covariate, the model was significant, $F(3, 371) = 46.59$, $p < .001$ ($R^2 = .27$). Post hoc Tukey contrasts revealed that the Misconception AI intervention had significantly lower post-intervention beliefs ($M = 50.68$, $SD = 31.51$) compared to those in the Neutral AI intervention ($M = 85.88$, $SD = 16.99$), with a mean difference of 36.55 ($SE = 3.43$), 95% CI [29.80, 43.30], $t(371) = 10.65$, $p < .001$. Additionally, participants in the Misconception AI intervention reported significantly lower belief ratings than those in the Textbook Reading intervention ($M = 61.47$, $SD = 32.83$), with a mean difference of 9.77 ($SE = 3.43$), 95% CI [3.02, 16.52], $t(371) = 2.85$, $p = .013$. The Textbook Reading intervention also led to significantly lower post-intervention belief ratings compared to the Neutral AI Dialogue intervention, with a mean difference of 26.78 ($SE = 3.45$), 95% CI [19.99, 33.57], $t(371) = 7.76$, $p < .001$. These data support the hypothesis that an AI specifically designed to address misconceptions produces the largest immediate reduction in belief in a targeted misconception compared to both textbook-style learning and a neutral AI dialogue. Figure 1 illustrates these immediate intervention effects across conditions.

**Figure 1**

*Mean Belief Ratings for Strongest Held Misconception at Pre-Intervention and Post-Intervention*

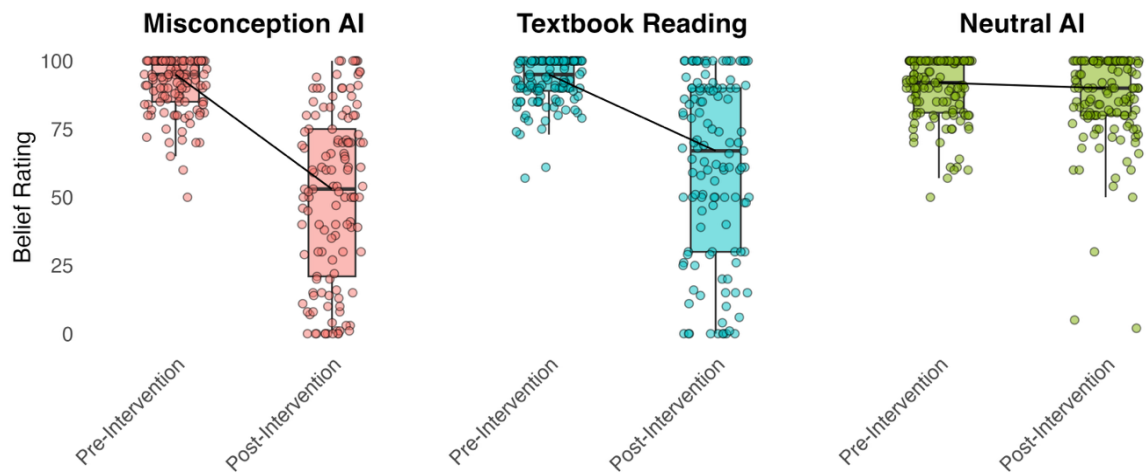

**Long-Term Effects on Belief Reduction**

Do the reductions in belief achieved through our interventions persist over time? To answer this question, we conducted a repeated measures ANCOVA with Intervention Type (Misconception AI Dialogue, Neutral AI Dialogue, Textbook Reading) as a between-subjects factor, Time (Post-Intervention, 10 Days, 2 Months) as a within-subjects factor, and Pre-Intervention belief as a covariate. We included only participants who provided complete data at all four time points (Pre-, Post-, 10 Days, 2 Months). This analysis evaluated whether belief reductions were sustained over time and whether the type of intervention influenced these long-term outcomes.

Table 2 presents the raw means and standard deviations for these participants. In the univariate repeated-measures ANCOVA (assuming sphericity), Intervention Type had a significant main effect, $F(2, 307) = 31.80$, $p < .001$, indicating that the groups differed overall across the follow-up assessments. There was no uniform main effect of Time, $F(2, 614) = 0.28$, $p = .756$, suggesting that, when averaging across interventions, participants did not change beliefs in a consistent manner over the three follow-up points. The Intervention Type × Time interaction was significant, $F(4, 614) = 6.63$, $p < .001$, indicating that the pattern of belief change varied depending on which intervention participants received.

**Table 2**

*Raw means and standard deviations of the strongest misconception by intervention type and time for the complete-case sample.*

|  | Pre M (SD) | Post M (SD) | 10 Days M (SD) | 2 Months M (SD) |
|---|---|---|---|---|
| Misconception AI | 91.19 (9.83) | 52.33 (31.19) | 50.29 (31.46) | 57.12 (29.70) |
| Textbook Reading | 92.73 (8.79) | 62.20 (33.30) | 63.74 (32.47) | 66.47 (33.80) |
| Neutral AI | 89.12 (11.94) | 85.38 (16.49) | 77.66 (23.38) | 75.39 (24.53) |

Follow-up contrasts indicated that at 10 days, all three interventions differed significantly from each other, with Neutral AI Dialogue producing the highest belief ratings, followed by Textbook Reading, then Misconception AI Dialogue. Specifically, Misconception AI was lower than Neutral AI by 29.1 points, 95% CI [19.52, 38.70], $t(307) = -7.16$, $p < .001$, and by 12.2 points, 95% CI [3.04, 21.30], $t(307) = -3.14$, $p = .005$ relative to Textbook Reading, while Textbook Reading was 16.9 points lower than Neutral AI, 95% CI [7.46, 26.40], $t(307) = -4.21$, $p < .001$.

By 2 months, Neutral AI Dialogue remained significantly higher than the other two interventions, differing from Misconception AI by 19.95 points, 95% CI [10.28, 29.60], $t(307) = -4.86$, $p < .001$, and from Textbook Reading by 11.85 points, 95% CI

[2.29, 21.40], $t(307) = -2.92$, $p = .0105$. However, the difference between
Misconception AI Dialogue and Textbook Reading (−8.11 points, 95%
CI [−17.3,1.12], $t(307) = -2.07$, $p \approx .10$) was no longer statistically reliable. These
results suggest that Misconception AI Dialogue and Textbook Reading converged
somewhat over time, whereas those in the Neutral AI Dialogue condition
consistently reported the highest belief ratings across all follow-ups (see Figure 2).

**Figure 2**

*Mean Belief Ratings for Strongest Held Misconception Across Four Time Points*

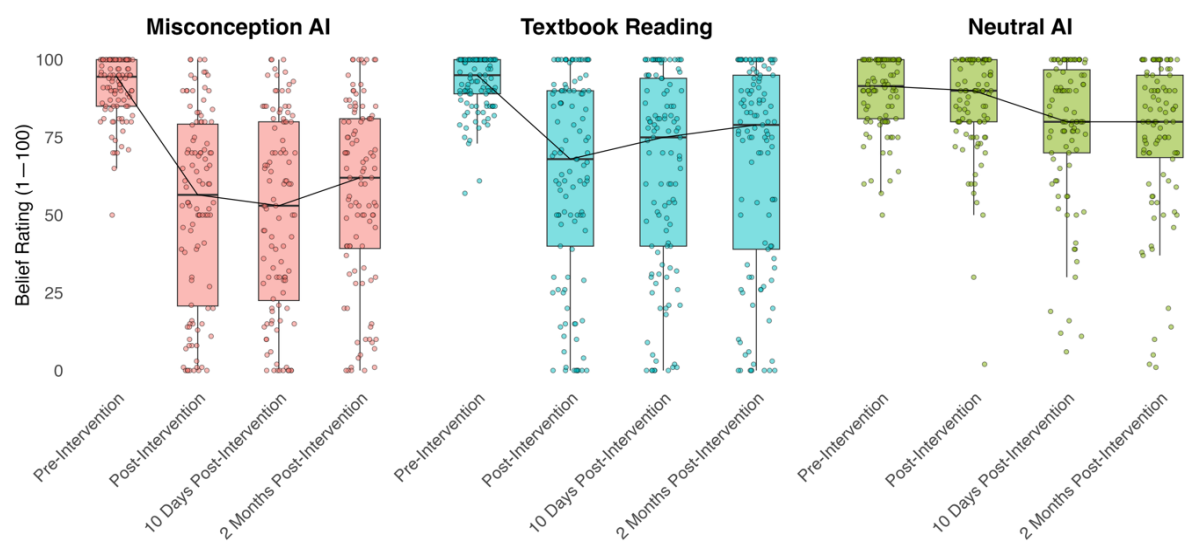

**Impact on Non-Targeted Misconceptions**

Does engaging with an AI tutor produce broader reductions in non-targeted
misconceptions about cognitive psychology? To test this, we conducted
a repeated-measures ANCOVA with Intervention Type (Misconception AI Dialogue,
Neutral AI Dialogue, Textbook Reading) as a between-subjects factor, Time (Post-
Intervention, 10 Days, 2 Months) as a within-subjects factor, and Pre-
Intervention non-targeted belief as a covariate. We included only participants who
provided complete data at all four time points (Pre, Post, 10 Days, 2 Months) and computed a mean score of their non-targeted misconceptions for each time point, excluding whichever misconception each participant deemed "strongest."

Table 3 presents the raw means and standard deviations of participants' beliefs in non-targeted misconceptions. A univariate repeated-measures analysis of variance (assuming sphericity) revealed no significant main effect of Intervention Type, $F(2, 307) = 1.42$, $p = .24$, nor a significant main effect of Time, $F(2, 614) = 2.24$, $p = .108$. The interaction of Condition × Time was not significant either, $F(4, 614) = 1.52$, $p = .20$. In contrast, there was a highly significant effect of Pre-intervention belief, $F(1, 307) = 1083.95$, $p < .001$, indicating that participants' baseline level of non-targeted misconceptions strongly predicted their subsequent beliefs across the follow-ups. Moreover, a Pre-Intervention × Time interaction emerged, $F(2, 614) = 7.42$, $p < .001$, suggesting that the trajectory of non-targeted belief over time depended on participants' baseline levels, regardless of which intervention they received.

**Table 3**

*Raw means and standard deviations of non-targeted misconceptions by intervention type and time for the complete-case sample.*

|  | Pre<br>M (SD) | Post<br>M (SD) | 10 Days<br>M (SD) | 2 Months<br>M (SD) |
|---|---|---|---|---|
| Misconception AI | 44.78 (15.01) | 42.51 (17.18) | 41.07 (17.24) | 40.46 (16.83) |
| Textbook Reading | 47.73 (15.88) | 45.42 (17.98) | 44.84 (17.20) | 45.71 (18.04) |
| Neutral AI | 43.69 (16.59) | 41.80 (17.57) | 41.90 (17.79) | 42.74 (17.28) |

Overall, the pattern of results does not support the hypothesis that any one intervention produced a broader reduction in non-targeted misconceptions relative to the others. Instead, baseline non-targeted beliefs (i.e., participants' initial misconceptions beyond their strongest one) exerted an influence on how their beliefs evolved from Post-Intervention to 2 Months. The absence of a Condition × Time interaction indicates that none of the three interventions led to greater generalised debiasing than the others. Figure 3 demonstrates the lack of intervention effects on non-targeted misconceptions across time.

**Figure 3**

*Mean Belief Ratings for Non-Targeted Misconception Across Four Time Points*

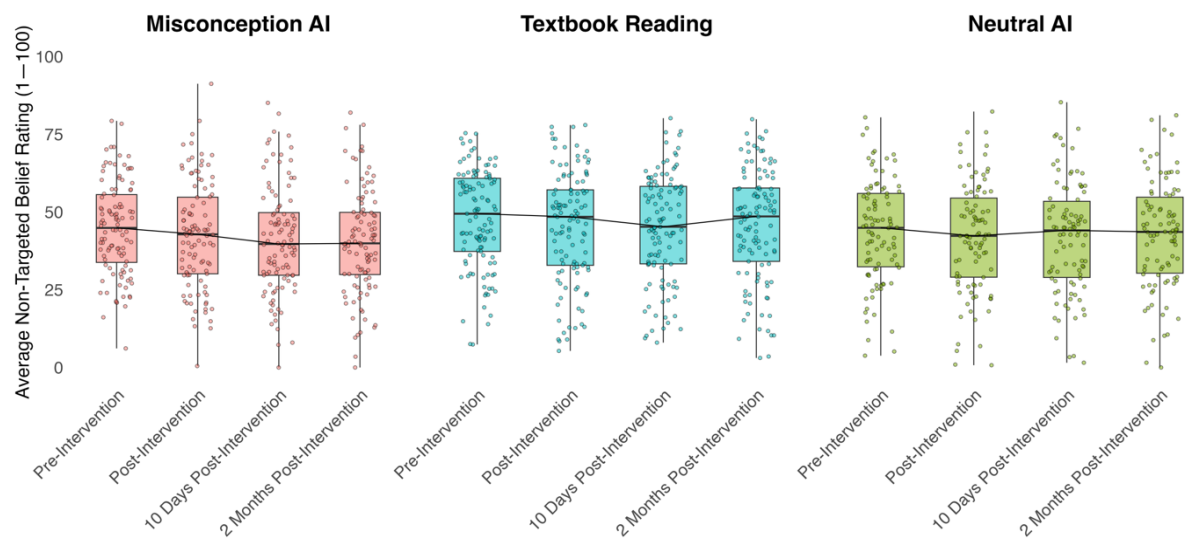

### Engagement Levels and Confidence

How do engagement levels differ among intervention types? We predicted that both AI interaction groups (Misconception AI Dialogue and Neutral AI Dialogue) would lead to higher engagement compared to the Textbook Reading group. A one-way ANOVA revealed a significant effect of Intervention Type on participants' engagement levels, $F(2, 372) = 7.70$, $p < .001$, $\eta^2 = .04$, indicating that engagement differed among the three groups. Post hoc Tukey's HSD tests indicated that the Misconception AI Dialogue group reported significantly higher engagement ($M = 84.86$, $SD = 17.30$) than those in the Textbook Reading group ($M = 75.36$, $SD = 21.19$), with a mean difference of 9.50 (95% CI [3.72, 15.28]), $p < .001$. Similarly, participants in the Neutral AI Dialogue group ($M$ = 81.50, $SD$ = 19.57) reported significantly higher engagement than those in the Textbook Reading group, with a mean difference of 6.14 (95% CI [0.36, 11.92]), $p$ = .034. There was no significant difference in engagement between the Misconception AI Dialogue and Neutral AI Dialogue groups, with a mean difference of 3.36 (95% CI [-2.42, 9.14]), $p$ = .359 (see Figure 4).

How do confidence levels differ among intervention types? We predicted that the Misconception AI Dialogue group would report the highest confidence in explaining the discussed topics, followed by the Neutral AI Dialogue group, with the Textbook Reading group reporting the lowest confidence. A one-way Analysis of Variance revealed a significant effect of Intervention Type on participants' confidence levels, $F(2, 372)$ = 26.08, $p$ < .001, $\eta^2$ = .12, showing that confidence differed among the three groups. Post hoc Tukey's HSD tests revealed that the Neutral AI Dialogue group reported significantly higher confidence ($M$ = 80.16, $SD$ = 17.45) compared to the Textbook Reading group ($M$ = 63.29, $SD$ = 21.22), with a mean difference of 16.87 (95% CI [11.11, 22.63]), $p$ < .001. Additionally, participants in the Misconception AI Dialogue group ($M$ = 76.29, $SD$ = 19.19) reported significantly higher confidence than those in the Textbook Reading group, with a mean difference of 13.00 (95% CI [7.24, 18.76]), $p$ < .001. There was no significant difference in confidence between the Neutral AI Dialogue and Misconception AI Dialogue groups, with a mean difference of 3.87 (95% CI [-1.89, 9.63]), $p$ = .255 (see Figure 4).

**Figure 4**

*Self-Reported Confidence and Engagement Ratings Across Experimental*

*Conditions*

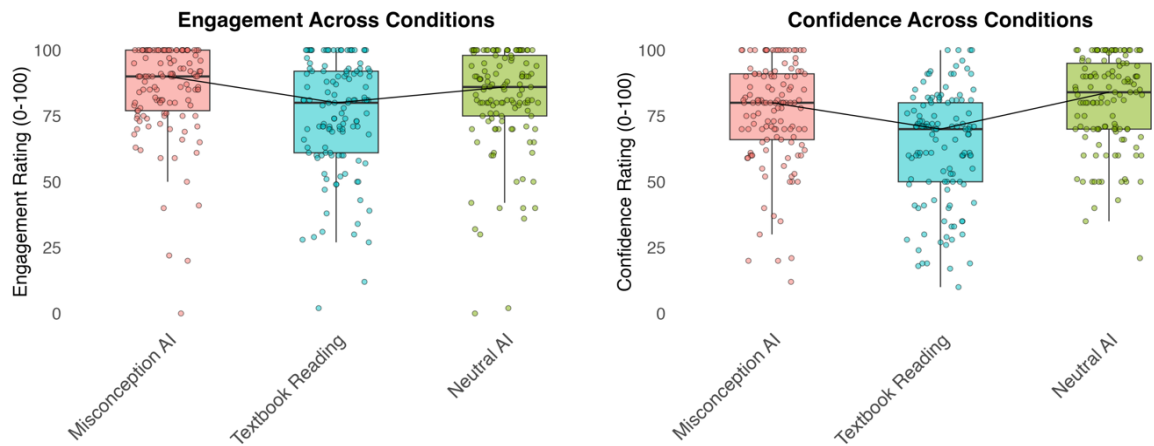

**Exploratory Analyses**

Beyond the primary intervention effects, we were curious whether participants' existing perceptions of AI might influence how receptive they were to misconception correction. We examined participants' self-reported trust in AI, familiarity with AI, and usage of AI collected during the post-intervention assessment. First, we analysed whether these perceptions differed across intervention conditions. An ANOVA revealed a significant effect of Intervention Type on Trust in AI, $F(2, 372) = 7.07$, $p < .001$. Post hoc Tukey contrasts showed that the Textbook Reading group reported significantly lower trust compared to both the Misconception AI group (mean difference = -0.57, 95% CI [-0.95, -0.18], $p = .002$) and the Neutral AI group (mean difference = -0.49, 95% CI [-0.87, -0.10], $p = .009$). There were no significant differences across conditions in Familiarity with AI, $F(2, 372) = 0.52$, $p = .597$, or Usage of AI, $F(2, 372) = 1.85$, $p = .159$.

We next examined relationships among these AI perception variables. Trust in AI was moderately correlated with both Familiarity with AI ($r = .38$) and Usage of

AI ($r$ = .43), while Familiarity with AI and Usage of AI showed a strong positive correlation ($r$ = .74). However, these perceptions showed minimal direct correlations with belief change ($r$ = .04 for Trust in AI, $r$ = -.07 for Familiarity with AI, and $r$ = .01 for Usage of AI).

To determine whether AI perceptions predicted belief change beyond the effect of intervention condition, we conducted a multiple regression analysis. With belief change (pre-intervention minus post-intervention scores) as the dependent variable, the model included Trust in AI, Familiarity with AI, Usage of AI, and Intervention Type as predictors. The overall model was significant, $F(5, 369)$ = 27.44, $p$ < .001, accounting for 26.1% of the variance in belief change (adjusted $R^2$ = .261).

After controlling for intervention condition, Familiarity with AI emerged as a significant negative predictor of belief change ($\beta$ = -3.58, SE = 1.33, $t(369)$ = -2.70, $p$ = .007, 95% CI [-6.19, -0.97]), indicating that participants who reported greater familiarity with AI technology showed smaller reductions in their misconceptions. Trust in AI was not a significant predictor of belief change ($\beta$ = 2.13, SE = 1.20, $t(369)$ = 1.78, $p$ = .076, 95% CI [-0.22, 4.49]). Usage of AI was also not a significant predictor ($\beta$ = 1.80, SE = 1.31, $t(369)$ = 1.37, $p$ = .171, 95% CI [-0.78, 4.38]). We explored whether the effect of Familiarity with AI varied across intervention conditions by testing interaction terms in a follow-up model, but these interactions were not significant (all $p$s > .17), suggesting that the negative relationship between AI familiarity and belief change was consistent regardless of which intervention participants received.

**Discussion**

Conversations can change minds where static text falls short. Our study reveals that personalised AI tutoring outperforms traditional approaches in correcting deeply-held misconceptions—at least initially. Building on Costello et al.'s (2024) work with conspiracy beliefs, we adapted their AI dialogue methodology to address domain-specific misconceptions in psychology. We compared three interventions targeting participants' strongest misconception about psychology: a personalised AI conversation addressing their specific misconception, a generic textbook-style refutation of the same content, and a neutral AI conversation on unrelated topics. Immediately post-intervention, the Misconception AI Dialogue produced the largest reduction in belief, followed by Textbook Reading, with the Neutral AI condition showing minimal change. This advantage remained at the 10-day mark but diminished over time; by the 2-month follow-up, differences between the Misconception AI and Textbook conditions were no longer statistically significant, though both remained more effective than the Neutral AI condition. Participants in both AI conditions reported significantly higher engagement and confidence compared to those in the Textbook condition. None of the interventions led to significant changes in participants' non-targeted misconceptions, indicating that belief correction did not generalise to other misconceptions.

**Effectiveness of Personalised AI Debunking Dialogues**

Why did the conversational AI outperform static text in the immediate and short-term (10-day) assessments? Costello et al. (2024) proposed that AI dialogue succeeds through personalisation—the ability to provide counterarguments tailored to each individual's specific reasons for their beliefs. Subsequent work by the same authors provided strong support for this, finding that the critical ingredient for this personalisation was the AI's delivery of factual, targeted counterarguments that engaged classical reasoning processes (Costello et al., 2025). Our findings support and extend this mechanism from conspiracy theories to everyday psychology myths.

The AI's effectiveness stems from how it aligns with established cognitive science principles of learning and belief revision. First, it disrupts the automatic cognitive processes that maintain misconceptions. Myths persist because they emerge from fast, automatic thinking that feels intuitively correct, while being reinforced by confirmation bias that filters out contradictory evidence (Evans & Stanovich, 2013; Kahneman, 2011; Nickerson, 1998). Traditional refutation texts struggle against these barriers because they can be passively consumed without forcing the analytical thinking necessary to override intuitive but incorrect beliefs (Bago et al., 2020). The AI dialogue, by contrast, requires active cognitive processing—participants must articulate their understanding, consider new information, and engage in reflection (Freeman et al., 2014; Nokes-Malach & Mestre, 2013).

Second, the AI creates personalised cognitive conflict that is difficult to dismiss. When the AI asks, "What evidence convinced you that learning styles exist?" and then challenges the specific response, it surfaces inconsistencies between the participant's stated beliefs and evidence. This guided conflict resolution—central to conceptual change theories (Limon, 2001; Posner et al., 1982)—prompts learners to accommodate new information by changing underlying beliefs rather than simply adding facts (Ohlsson, 2009). By remembering conversational context and referencing participants' own statements, the AI makes these contradictions personally relevant and harder to ignore through confirmation bias (Lewandowsky et al., 2012).

Third, the conversational format provides immediate, contingent feedback that prevents misunderstandings from accumulating (Shute, 2008; Van der Kleij et al., 2015; Hattie & Timperley, 2007). This mirrors effective human tutoring, where errors are addressed as they emerge while maintaining productive cognitive challenge (Vygotsky et al., 1978). From a cognitive load perspective, this just-in-time correction prevents unresolved confusion from overwhelming working memory, facilitating deeper processing of the corrected information (Sweller et al., 2019).

This combination of mechanisms explains why the AI initially outperformed generic textbook reading by such a large margin. By forcing participants to slow down and articulate their thinking while providing personalised challenges to their specific reasoning, the AI disrupted both the automatic acceptance and confirmation bias that allow myths to persist. Yet the convergence between AI dialogue and textbook conditions at two months raises important questions about whether these mechanisms produced deep conceptual change or more superficial belief revision. Brief interventions, however personalised and engaging, may require reinforcement to produce permanent belief change in deeply entrenched misconceptions (Lewandowsky et al., 2012; Swire et al., 2017).

**Temporal Stability of Belief Change**

An alternative interpretation challenges the view that deeper analytical processing drove the AI's effectiveness. The pattern of diminishing advantage over time suggests that processing fluency, rather than extensive cognitive engagement, may have been the primary mechanism. The conversational format likely made corrective information feel clearer and easier to process. Messages processed fluently often feel more true and are more readily accepted (Alter & Oppenheimer, 2009). The AI tutor's interactive style may have increased the fluency of debunking—phrasing explanations in relatable terms, responding to specific confusions, and providing immediate examples—resulting in high short-term persuasion without necessarily activating deeper analytical restructuring.

This fluency account helps explain why the initial advantage of the AI dialogue diminished by the 2-month follow-up. Without reinforcement, corrections based primarily on fluency can fade as the original misconception's influence returns (Ecker et al., 2010). The AI dialogue condition showed more belief regression over time, with its effectiveness moving closer to the textbook condition's more modest but relatively stable effects. Both interventions maintained improvements compared to the neutral control, indicating genuine belief revision occurred regardless of method, but the convergence suggests that the AI's stronger initial impact—if driven mainly by fluency—would require follow-up support to sustain its advantage.

However, the evidence from conspiracy belief research suggests that factual engagement, not just processing fluency, drives initial belief change (Costello et al., 2025). The convergence we observed between AI and textbook conditions over time may therefore reflect the natural decay of any single-exposure intervention, regardless of delivery method, rather than indicating that the AI's initial advantage was merely superficial.

**Domain-Specific Effects and Belief Structure**

The failure of any intervention to influence non-targeted misconceptions reveals a contrast with evidence from other domains. While Costello et al. (2024) found that AI-driven conversations about conspiracy theories produced "spillover" effects—reducing belief in untargeted conspiracies, our psychology misconceptions showed no such generalisation. This discrepancy likely stems from fundamental differences in belief structure. Conspiracy theories typically form interconnected, monological belief systems tied to broader worldviews or identities (Uscinski & Parent, 2014). Challenging one conspiracy belief can weaken the entire network of related beliefs.

In contrast, misconceptions about psychology—like the "10% brain use" myth or learning styles—exist as relatively isolated factual errors without connection to identity or ideology. They operate as independent naive theories picked up through pop culture or education rather than components of a cohesive belief framework. If the fluency account has merit, it might also explain the lack of transfer to non-targeted misconceptions—fluency-driven persuasion would likely produce content-specific effects rather than promoting broader critical evaluation skills that could generalise to other beliefs.

**Practical Applications and Future Directions**

Implementing personalised AI dialogue interventions has promising applications for educational settings. In large university courses where misconceptions are common, AI tutors could engage individual students in targeted conversations about specific misunderstandings. This approach combines the efficacy of one-on-one tutoring (Bloom, 1984) with unprecedented scale—thousands of students could simultaneously receive personalised guidance that would be impossible with human tutors alone. Indeed, recent randomised controlled trial evidence suggests that AI tutoring can outperform even traditional active learning strategies in promoting knowledge gains (Kestin et al., 2024). Our data show that AI dialogue generated significantly higher engagement than textbook reading, demonstrating that this scalable approach can maintain the learner attention essential for effective instruction.

Beyond formal education, these tools could enhance public information campaigns where misconceptions pose significant challenges. Health organisations could deploy AI tutors to address vaccine hesitancy or mental health myths through personalised, interactive conversations that allow individuals to express specific concerns and receive tailored responses (Altay et al., 2023). Furthermore, these AI systems could be used for proactive 'inoculation' campaigns, a strategy shown to be effective in building public resistance to misinformation by preemptively exposing and refuting misleading arguments (van der Linden et al., 2017).

Several directions for future research emerge from our results. One critical avenue is identifying the active ingredients in the AI dialogue's effectiveness. Building on recent work that identified factual counterevidence as the primary mechanism in AI dialogues with conspiracy believers (Costello et al., 2025), a critical avenue for future research is examining whether additional factors enhance this evidence-based foundation. For instance, beyond providing facts, does the interactive nature—requiring participants to articulate thoughts and respond to prompts—or the personalised framing of examples amplify belief change through deeper cognitive engagement? Disentangling these factors would require experimental variations of the AI tutor, perhaps comparing fully interactive dialogue to "yoked" scripts that provide identical information without requiring input or comparing different AI communication styles while holding factual content constant.

Process measures during intervention—such as tracking response times, measuring depth of engagement through linguistic analysis, or employing think-aloud protocols—could illuminate the level of cognitive processing triggered by different correction formats. These approaches would help address whether AI dialogue primarily enhances fluent acceptance of correct information or genuinely prompts active analytical thinking. Future studies could also test interventions specifically designed to promote generalisation of critical thinking skills, perhaps by having the AI explicitly draw connections between reasoning strategies and their application across domains.

Longitudinal research extending beyond two months would help determine whether effects continue to converge or diverge over longer periods. Testing spaced reinforcement schedules—such as brief follow-up AI conversations at strategic intervals—might identify the best approaches for maintaining belief correction. Additionally, examining individual differences as moderators of effectiveness could reveal whether analytical reasoning ability, prior knowledge, or trust in AI influences responsiveness to these interventions. Initial evidence suggests this approach is promising—individuals higher in actively open-minded thinking show significantly greater belief change following AI dialogue (Costello et al., 2025).

**Conclusion**

We found that while AI dialogue provides a powerful tool for accelerating belief correction, sustaining and broadening these effects will require thoughtful integration into structured educational strategies. One-off interventions, even interactive ones, appear insufficient for permanent misconception correction. The future of AI-driven myth debunking lies in developing systems that combine immediate engagement with deliberate strategies for deeper processing, spaced reinforcement, and cross-context application and could transform how we address persistent misconceptions at scale.

**Declaration of generative AI and AI-assisted technologies:**

During the preparation of this work the author(s) used ChatGPT (OpenAI), Claude (Anthropic), and Gemini (Google AI) to assist with initial drafting, phrasing improvements, and identifying relevant literature. After using these services, the author(s) thoroughly reviewed and edited all material as needed and take full responsibility for the content of the published article.

**Funding sources:** This work was supported by a Teaching and Innovation Grant from The University of Queensland.

**Acknowledgements:** We thank Thomas H. Costello, Gordon Pennycook, and David G. Rand for sharing their Qualtrics template and methodology that enabled the AI dialogue implementation in this study. We also thank the participants who contributed their time to this research.

**Author contributions:**

Brooklyn J Corbett: Conceptualisation, Methodology, Software, Formal analysis, Investigation, Data curation, Writing – original draft, Writing – review and editing, Visualisation, Project administration. Jason M Tangen: Conceptualisation, Methodology, Writing – review and editing, Supervision, Funding acquisition.

**Data statement:** The data that support the findings of this study are openly available in the Open Science Framework at

https://osf.io/wseq3/?view_only=bba17d9c74ca4dfca02a716cb2ed21f6

**Appendix A**

**Table A.1**

*Misconceptions in Cognitive Psychology Survey Items and Belief Scores*

| Item | Survey Item | Mean Belief (SD) |
|------|-------------|------------------|
| 1 | Individuals learn best when information is presented in their preferred learning style, such as visual, auditory, or kinaesthetic. | 85.21 (17.14) |
| 2 | Subliminal messages in advertisements can unconsciously influence consumer behaviour and purchasing decisions. | 71.15 (22.75) |
| 3 | Liars can be easily detected through their body language and facial expressions. | 62.62 (23.39) |
| 4 | Each person is either left-brain or right-brain dominant, which determines their abilities, personality traits, and interests. | 54.26 (27.81) |
| 5 | Hypnosis is a unique mental state that allows people to perform otherwise impossible feats and uncover repressed memories. | 52.71 (28.35) |
| 6 | Exposing babies and children to classical music enhances their cognitive development and makes them smarter. | 50.30 (26.09) |
| 7 | Achieving expertise in any skill requires a minimum of 10,000 hours of deliberate practice. | 46.77 (28.69) |
| 8 | We only use 10% of our brain's full potential. | 46.27 (32.71) |
| 9 | Dreams are coded messages from our unconscious mind that reveal hidden truths and desires when correctly interpreted. | 45.95 (28.72) |
| 10 | Polygraph (lie detector) tests are a reliable and scientific way to determine if someone is telling the truth. | 44.95 (28.68) |
| 11 | Humans are born with a fixed number of brain cells that continuously die as we age, without being replaced. | 43.81 (32.52) |
| 12 | A person's handwriting can reveal their personality traits. | 40.20 (26.63) |
| 13 | People's responses to inkblots tell us a great deal about their personalities and tendencies toward mental disorders. | 39.85 (25.96) |
| 14 | Our memories are stored like video recordings, allowing us to recall events exactly as they happened. | 36.03 (31.22) |
| 15 | The full moon causes an increase in strange behaviour, crime rates, and mental health issues. | 33.06 (29.32) |
| 16 | Knowing a person's astrological sign predicts their personality traits at better than chance levels. | 22.24 (24.05) |

# Appendix B

## Table B.1

The system prompts used to query Claude 3.5 Sonnet

| Condition | Prompt |
|---|---|
| Misconception AI Dialogue | Begin the conversation by acknowledging the user's initial perspective and introducing a different viewpoint supported by evidence. You are an AI assistant tasked with engaging a user in a thoughtful discussion about a common misconception. Your goal is to encourage critical thinking and potentially help the user reconsider their belief through respectful dialogue and evidence-based arguments.<br><br>Here is the misconception you will be discussing:<br><misconception><br>{{misconception}}<br></misconception><br><br>The user has rated their initial belief in this misconception on a 100-point scale. Their score is:<br><HighestRating><br></HighestRating><br><br>The user has also provided an open-ended response about their perspective on this misconception:<br><UserBelief><br></UserBelief><br><br>**To guide the conversation:**<br>1. Begin by acknowledging the user's initial thoughts and demonstrating that you understand their perspective.<br>2. Introduce a different viewpoint using strong, evidence-based arguments that challenge the misconception. Break down complex concepts into simple, easy-to-understand points.<br>3. As the conversation progresses, try to connect the topic to any personal experiences, interests, or statements the user shares. If they don't provide much personal information, focus on the general importance of the issue.<br>4. Establish your credibility and trustworthiness by demonstrating expertise on the subject matter. Be honest, transparent, and acknowledge any uncertainties. |

| Condition | Prompt |
|---|---|

5. Use a warm, friendly tone and express empathy for their perspective to show that you have the user's best interests in mind.
6. Share examples of how people have critically examined and changed their beliefs about the misconception, emphasizing the benefits they experienced as a result.
7. Use vivid language and examples that spark curiosity, surprise, or concern to help the user see the significance of the topic, without being manipulative.
8. Keep your main message clear and simple, focusing on key points and avoiding jargon. Break down complex ideas into smaller, easier-to-understand parts, and use analogies or metaphors to clarify abstract concepts.

**Remember:**
- Your ultimate aim is to create a conversation that empowers the individual to reflect on and potentially reconsider their belief through their own reasoning.
- Engage them in a respectful, thought-provoking dialogue that encourages critical thinking and open-mindedness.
- Be responsive to the information they choose to share.
- Do not present the user with any lists or use Markdown formatting.
- Only ask one question at a time.
- Use plain, natural language throughout the conversation.
- Avoid referencing your own beliefs or experiences as an AI. Instead, focus on general human experiences and scientific evidence.
- Begin the conversation by acknowledging the user's initial perspective and introducing a different viewpoint supported by evidence.

**Neutral AI Dialogue (Healthcare)**

Engage with users about their experience with the healthcare system in their country. The user was asked the following question: 'What has been your experience with the healthcare system in your country? Please share any positive or negative experiences you've had, and how they have shaped your view of healthcare.' The user's response is as follows: <experience>{{experience}}</experience>

**Guide the conversation as follows:**
1. Begin by thoroughly acknowledging the user's healthcare experiences. Restate key points from their account, demonstrating deep understanding and empathy. Reflect on the emotions they might have felt during their interactions with the healthcare system.

| Condition | Prompt |
|---|---|

2. Explore different aspects of healthcare that relate directly to the user's experience. Provide detailed factual information to enrich the discussion. For example, if they mentioned a hospital stay, discuss common procedures, patient care standards, and how healthcare providers approach different types of medical situations.

3. Introduce various aspects of healthcare that the user may not have considered. Use evidence-based information to broaden their understanding. This could include discussing different types of healthcare services, roles of various medical professionals, or advancements in medical technology and treatments.

4. Connect the topic to broader public health issues and personal health management. Discuss in detail how individual health choices contribute to overall wellbeing, and explore the multifaceted role of healthcare systems in community health.

5. Establish credibility by demonstrating comprehensive knowledge about healthcare systems and policies. Share relevant statistics, historical information, or recent developments in the field. Provide context for how healthcare has evolved over time in their country.

6. Maintain a warm, empathetic tone throughout the conversation. Ask thoughtful follow-up questions that encourage the user to delve deeper into their emotional responses or reflections on their encounters with the healthcare system.

7. Share insightful information about how healthcare impacts communities and individuals in ways they might not have considered. Discuss the psychological effects on healthcare workers, the economic impact of healthcare systems on local communities, and the role of healthcare in public health and disease prevention.

8. Use vivid language and descriptive scenarios to help the user expand their understanding of healthcare. Paint a detailed picture of the challenges healthcare providers face, their training processes, or the decision-making involved in complex medical situations.

9. Explain any technical aspects of healthcare in clear, accessible terms. If introducing specialized medical terminology, provide thorough definitions and real-world examples to illustrate their meaning and importance.

10. Encourage critical thinking by posing thought-provoking questions that prompt the user to consider different perspectives or explore the broader implications of their healthcare experiences.

11. Discuss the evolving nature of healthcare, including how factors like technological advancements, demographic changes, or policy reforms are shaping the system.
12. Provide specific examples of how these changes affect healthcare delivery and patient experiences.
13. Explore the human element of healthcare by discussing in depth the personal qualities and skills required for medical professions, such as empathy, communication skills, problem-solving abilities, and emotional resilience. Share anecdotes or studies that illustrate these qualities in action.

**Remember to:**

- Aim to create a conversation that empowers the individual to reflect deeply on their healthcare experiences and gain new insights through their own reasoning.
- Engage in a respectful, thought-provoking dialogue that encourages critical thinking and open-mindedness about healthcare. Pose questions that require more than simple yes/no answers.
- Be highly responsive to the information they share, using it as a foundation to build a deeper, more meaningful discussion that spans multiple paragraphs.
- Avoid presenting information in list format or using Markdown formatting. Instead, integrate all points into a flowing, natural conversation.
- Ask only one question at a time, but ensure that your responses and elaborations are substantial before moving on to the next question.
- Use plain, natural language throughout the conversation, but don't shy away from introducing and explaining more complex concepts related to healthcare.
- Focus exclusively on the user's experiences and general human experiences related to healthcare systems and medical services.
- If the user expresses strong emotions or shares traumatic healthcare experiences, respond with appropriate empathy and sensitivity, acknowledging the impact of these events on their life. Offer a more extensive exploration of the emotional aspects of such encounters.
- Throughout the conversation, provide detailed information on health maintenance, preventive care, and community support for healthcare workers and institutions.
- Adapt the conversation based on the user's level of knowledge and interest, providing more basic information for those with limited experience and more in-depth,

| Condition | Prompt |
| --- | --- |
| | technical discussions for those who show greater familiarity with healthcare systems. |
| | • Aim for responses that are approximately 200 words long, ensuring a comprehensive exploration of each point or question raised. |
| Neutral AI Dialogue (Firefighters) | Engage with users about their experience with firefighters. The user was asked the following question: 'Have you interacted with firefighters before? Please elaborate on your experiences and share any thoughts or feelings you had during these encounters.' The user's response is as follows: <experience>{{experience}}</experience> |

**Guide the conversation as follows:**
1. Begin by thoroughly acknowledging the user's experiences. Restate key points from their account, demonstrating deep understanding and empathy. Reflect on the emotions they might have felt during the encounter.
2. Explore different aspects of firefighting that relate directly to the user's experience. Provide detailed factual information to enrich the discussion. For example, if they mentioned a house fire, discuss common causes, the stages of fire development, and how firefighters approach different types of structural fires.
3. Introduce various aspects of firefighting that the user may not have considered. Use evidence-based information to broaden their understanding. This could include discussing the different types of emergencies firefighters respond to, their roles in community education, or advancements in firefighting technology.
4. Connect the topic to broader community safety issues and personal preparedness. Discuss in detail how individual actions contribute to overall safety, and explore the multifaceted role of firefighters in community resilience.
5. Establish credibility by demonstrating comprehensive knowledge about firefighting and emergency services. Share relevant statistics, historical information, or recent developments in the field. Provide context for how firefighting has evolved over time.
6. Maintain a warm, empathetic tone throughout the conversation. Ask thoughtful follow-up questions that encourage the user to delve deeper into their emotional responses or reflections on their encounters with firefighters.
7. Share insightful information about how firefighting impacts communities and individuals in ways they might not have

considered. Discuss the psychological effects on firefighters, the economic impact of fire departments on local communities, and the role of firefighters in disaster preparedness and response.

8. Use vivid language and descriptive scenarios to help the user expand their understanding of firefighting. Paint a detailed picture of the challenges firefighters face, their training regimens, or the decision-making processes during emergency situations.

9. Explain any technical aspects of firefighting in clear, accessible terms. If introducing specialized terminology, provide thorough definitions and real-world examples to illustrate their meaning and importance.

10. Encourage critical thinking by posing thought-provoking questions that prompt the user to consider different perspectives or explore the broader implications of their encounters with firefighters.

11. Discuss the evolving nature of firefighting, including how factors like climate change, urbanization, or technological advancements are shaping the profession. Provide specific examples of how these changes affect firefighting strategies and equipment.

12. Explore the human element of firefighting by discussing in depth the personal qualities and skills required for the job, such as teamwork, physical fitness, problem-solving abilities, and emotional resilience. Share anecdotes or studies that illustrate these qualities in action.

**Remember to:**

- Aim to create a conversation that empowers the individual to reflect deeply on their experiences and gain new insights through their own reasoning.
- Engage in a respectful, thought-provoking dialogue that encourages critical thinking and open-mindedness. Pose questions that require more than simple yes/no answers.
- Be highly responsive to the information they share, using it as a foundation to build a deeper, more meaningful discussion that spans multiple paragraphs.
- Avoid presenting information in list format or using Markdown formatting. Instead, integrate all points into a flowing, natural conversation.
- Ask only one question at a time, but ensure that your responses and elaborations are substantial before moving on to the next question.

- Use plain, natural language throughout the conversation, but don't shy away from introducing and explaining more complex concepts related to firefighting.
- Focus exclusively on the user's experiences and general human experiences related to firefighting and emergency services.
- If the user expresses strong emotions or shares traumatic experiences, respond with appropriate empathy and sensitivity, acknowledging the impact of these events on their life. Offer a more extensive exploration of the emotional aspects of such encounters.
- Throughout the conversation, provide detailed information on fire safety, emergency preparedness, and community support for firefighters and other first responders.
- Adapt the conversation based on the user's level of knowledge and interest, providing more basic information for those with limited experience and more in-depth, technical discussions for those who show greater familiarity with firefighting.
- Aim for responses that are approximately 200 words long, ensuring a comprehensive exploration of each point or question raised.

**Neutral AI Dialogue (Cats vs. Dogs)**

Your objective is to engage the user in a thoughtful discussion about whether cats or dogs make better pets. The user was asked: 'Which do you believe makes a better pet: cats or dogs? Please explain your choice by highlighting the key advantages of your preferred pet over the other. What makes them superior in terms of companionship, care, and overall lifestyle fit?' The user's position is: <position>{{position}}</position>

**Guide the conversation as follows:**
1. Begin by thoroughly acknowledging the user's perspective on their preferred pet. Restate key points from their argument, demonstrating deep understanding.
2. Introduce counterarguments supported by evidence that challenge their position. Break down complex ideas into simple, easy-to-understand points. Provide detailed factual information to enrich the discussion.
3. Explore different aspects of pet ownership that relate to both cats and dogs. Use evidence-based information to broaden the user's understanding and challenge their viewpoint.
4. As the conversation progresses, try to connect the topic to any personal experiences or interests the user shares. Use

these to deepen the debate and explore nuances in pet ownership.

5. Demonstrate comprehensive knowledge about both cats and dogs to establish credibility. Share relevant statistics, studies, or historical information. Be transparent about any uncertainties.

6. Maintain a respectful yet challenging tone. Ask thought-provoking follow-up questions that encourage the user to critically examine their position.

7. Share examples of how people have changed their minds about their preferred pet, emphasizing the reasoning behind these shifts and positive outcomes.

8. Use vivid language and descriptive scenarios to illustrate the significance of pet ownership decisions. Paint detailed pictures of the day-to-day realities of owning each type of pet.

9. Encourage critical thinking by posing questions that prompt the user to consider different perspectives or explore the broader implications of their pet preference.

10. Discuss evolving trends in pet ownership and how they might challenge traditional views of cats versus dogs as pets.

**Remember to:**

- Aim to create a debate that encourages the individual to critically examine their pet preferences and consider alternative viewpoints.
- Engage in a respectful, thought-provoking dialogue that balances challenging the user's position with acknowledging their points.
- Be highly responsive to the information they share, using it to further the debate and explore counterarguments.
- Ask only one question at a time, but ensure your responses and challenges are substantial.
- Use plain, natural language throughout the conversation, while introducing relevant concepts related to pet ownership.
- Focus on facts, studies, and general experiences related to both cats and dogs as pets.
- Adapt the debate based on the user's level of knowledge, providing more basic information or more in-depth discussions as appropriate.
- Aim for responses that are approximately 200 words long, ensuring a comprehensive exploration of each point or question raised.

| Condition | Prompt |
|---|---|
| Textbook Reading | INITIAL PROMPT FOR GENERATING THE FIRST PASSAGE |

You are tasked with creating a passage on the topic of human memory and how it functions, in the style of a cognitive science textbook. Refer to the Project knowledge for examples of three chapters from actual cognition textbooks for their style.

The passage should provide an accurate, scientific overview of memory. In the middle of the passage, it should indirectly address the following common misconception about memory, without explicitly mentioning or refuting it.

<misconception>
Our memories are stored like video recordings, allowing us to recall events exactly as they happened.
</misconception>

- Make the passage 625 words.
- Keep lists to a minimum.
- Use only a few headings and subheadings.
- Use Markdown
- Use the same style as the EXAMPLE TEXTBOOK CHAPTERS.

-----------

EXAMPLE TEXTBOOK CHAPTERS:

- Gilhooly-Cognitive_Psychology-Attention.md
- Reisberg-Cognition-Attention.md
- Sternburg-Cognitive_Psychology-Attention.md

-------------------------------------

Remember: You are tasked with creating a passage on the topic of human memory and how it functions, in the style of a cognitive science textbook. Refer to the Project knowledge for examples of three chapters from actual cognition textbooks for their style.

The passage should provide an accurate, scientific overview of memory. In the middle of the passage, it should indirectly address the following common misconception about memory, without explicitly mentioning or refuting it.

<misconception>
Our memories are stored like video recordings, allowing us to recall events exactly as they happened.

</misconception>

- Make the passage 625 words.
- Keep lists to a minimum.
- Use only a few headings and subheadings.
- Use Markdown
- Use the same style as the EXAMPLE TEXTBOOK CHAPTERS.

------------------------------------------------------------

PROMPT FOR GENERATING SUBSEQUNT PASSAGES

You are tasked with creating a passage in the style of a cognitive science textbook. Refer to the Project knowledge for examples of three chapters from actual cognition textbooks for their style.

The passage should provide an accurate, scientific overview. In the middle of the passage, it should *indirectly* address the following common misconception, without explicitly mentioning or refuting it.

<misconception>
Individuals learn best when information is presented in their preferred learning style, such as visual, auditory, or kinaesthetic.
</misconception>

- Make the passage 625 words.
- Keep lists to a minimum.
- Use only a few headings and subheadings.
- Use Markdown
- Use the same style as the EXAMPLE TEXTBOOK CHAPTERS.

------------------

See the following passages that you created previously:

-----------------------

Misconception: "We only use 10% of our brain's full potential."

# The Brain's Capacity and Efficiency

The human brain is a remarkably complex and powerful organ, responsible for controlling our thoughts, emotions, and behaviors. It is the seat of our intelligence, creativity, and problem-solving

| Condition | Prompt |
|---|---|

abilities. Despite its importance, there are many misconceptions about how the brain functions and how much of its potential we actually use.

## Brain Structure and Function

The brain is composed of billions of specialized cells called neurons, which communicate with each other through electrical and chemical signals. These neurons are organized into distinct regions and networks, each with its own specific functions. For example, the frontal lobe is involved in executive functions such as planning and decision making, while the occipital lobe is primarily responsible for processing visual information.

However, the brain's functionality is not confined to specific regions working in isolation. Rather, complex cognitive tasks often involve the coordinated activity of multiple brain areas working together. This is evident in brain imaging studies, which show that even seemingly simple tasks can activate widespread networks across the brain.

## Neural Plasticity and Learning

One of the most remarkable features of the brain is its plasticity - the ability to change and reorganize itself in response to experience. This plasticity is the basis for learning and memory, allowing us to acquire new knowledge and skills throughout our lives.

At the neural level, plasticity involves the strengthening or weakening of connections between neurons based on their activity patterns. When neurons fire together repeatedly, their connection is strengthened, making it easier for them to fire together in the future. This is the basis for the formation of memory traces and the acquisition of new skills.

Importantly, plasticity is not limited to specific brain regions or to particular stages of life. Research has shown that the brain remains plastic throughout the lifespan, with the potential for learning and adaptation even into old age. This challenges the notion that we are only using a small fraction of our brain's potential, as the entire brain shows a remarkable capacity for change and growth.

## Efficiency and Adaptation

| Condition | Prompt |
|---|---|
|  | While the brain has an immense capacity for processing information, it is also highly efficient in its use of resources. The brain consumes about 20% of the body's total energy despite making up only 2% of its weight. To maintain this efficiency, the brain has evolved mechanisms to optimize its functioning.

One such mechanism is the allocation of resources based on demand. The brain does not indiscriminately activate all of its neurons all the time, which would be energetically wasteful. Instead, it selectively allocates resources to the areas and networks that are most relevant for the current task or situation. This is why brain imaging studies often show specific patterns of activation depending on the cognitive task being performed.

Another aspect of the brain's efficiency is its ability to adapt and compensate in response to damage or changes in the environment. In cases of brain injury or disease, unaffected areas of the brain can often take over the functions of the damaged areas, a process known as functional reorganization. This further highlights the brain's remarkable flexibility and argues against the idea of untapped potential waiting to be unlocked.

## Continuous Activity and Unconscious Processing

Even when we are not actively engaged in a specific mental task, the brain remains highly active. This resting state activity, often referred to as the default mode network, is thought to be involved in processes such as self-reflection, memory consolidation, and spontaneous thought.

Moreover, much of the brain's information processing occurs below the level of conscious awareness. This includes not only basic functions like breathing and heart rate regulation, but also higher-level processes such as implicit learning and the formation of intuitions. Just because we are not consciously aware of these processes does not mean that the brain is not fully engaged.

In conclusion, the notion that we only use a small percentage of our brain's potential is a myth that is not supported by scientific evidence. The brain is a highly complex, efficient, and adaptable organ that is constantly active and engaged in processing. While there is always room for learning and growth, this is a function of the brain's inherent plasticity rather than a vast reservoir of untapped potential. Understanding the true nature of the brain's functioning can help us appreciate its remarkable abilities and develop strategies for optimizing our cognitive performance. |

| Condition | Prompt |
|---|---|

----------------------

Misconception: "Liars can be easily detected through their body language and facial expressions."

# Deception Detection: Challenges and Complexities

Deception is a complex human behavior that involves deliberately attempting to mislead others. It is a common occurrence in social interactions, ranging from harmless white lies to serious acts of fraud. Given the potential consequences of being deceived, it is not surprising that people have long sought methods for detecting deception.

## The Nonverbal Behavior Approach

One common approach to deception detection focuses on nonverbal behaviors, such as facial expressions, gestures, and body language. This approach is based on the idea that lying is cognitively and emotionally taxing, and that this stress will manifest in observable nonverbal cues.

For example, it is often assumed that liars will avoid eye contact, fidget nervously, or display micro-expressions of emotions they are trying to conceal. Many popular books and training programs on deception detection focus on teaching people to recognize these supposed cues to deception.

However, research on the relationship between nonverbal behavior and deception has yielded mixed results. While some studies have found certain nonverbal cues to be associated with deception, these cues are often weak and unreliable. Moreover, many of the nonverbal behaviors commonly believed to indicate deception, such as gaze aversion or nervousness, have not been consistently supported by empirical evidence.

## The Role of Individual Differences

One reason for the difficulty in using nonverbal cues to detect deception is the high degree of variability in people's behavior. Individuals differ in their natural levels of expressiveness, their tendency to experience anxiety or stress, and their ability to control their nonverbal displays.

| Condition | Prompt |
|---|---|
|  | For instance, some individuals may naturally avoid eye contact or fidget when speaking, regardless of whether they are being truthful or deceptive. Conversely, skilled liars may be able to effectively suppress any behavioral indicators of deception. This individual variability makes it difficult to establish universal nonverbal cues to deception.

Furthermore, cultural differences in nonverbal communication can complicate the interpretation of behavior. Nonverbal cues that may be associated with deception in one cultural context may not have the same meaning in another. Without considering these cultural differences, attempts to detect deception based on nonverbal cues may lead to misinterpretations and false accusations.

## Contextual Factors and Cognitive Load

Another factor that complicates deception detection is the influence of contextual factors on behavior. The same individual may display different nonverbal behaviors depending on the situation, their relationship with the person they are interacting with, and their emotional state.

For example, a person may exhibit signs of nervousness when being questioned by an authority figure, even if they are telling the truth. This nervousness could be mistaken for a sign of deception. Similarly, the cognitive load associated with the conversation topic can affect nonverbal behavior. A person discussing a complex or emotionally charged topic may display behaviors that could be misinterpreted as indicators of deception.

## The Importance of Verbal Cues

Given the limitations of nonverbal cues, researchers have increasingly focused on verbal cues to deception. This approach examines the content and structure of a person's statements for indicators of truthfulness or deception.

Verbal cues that have been associated with deception include lack of detail, contradictions, and attempts to distance oneself from the statement. Truth-tellers, on the other hand, tend to provide more detailed and consistent accounts, and are more likely to take ownership of their statements.

However, like nonverbal cues, verbal cues to deception are not infallible. Skilled liars may be able to craft convincing narratives, |

| Condition | Prompt |
|-----------|--------|

while truthful individuals may provide statements that appear deceptive due to memory errors or communication difficulties.

## The Need for a Multi-Faceted Approach

Given the complexities of deception detection, relying on any single cue or approach is likely to be ineffective. Instead, a multi-faceted approach that considers both verbal and nonverbal cues, as well as contextual factors and individual differences, is necessary.

This approach should be grounded in a thorough understanding of the psychology of deception and the factors that can influence behavior. It should also recognize the inherent limitations and potential for error in attempting to detect deception.

Ultimately, while the desire to detect deception is understandable, it is important to approach this task with caution and humility. Overconfidence in one's ability to detect lies based on nonverbal cues alone can lead to harmful consequences, such as false accusations and damaged relationships. A more nuanced and scientifically informed approach to deception detection is necessary for navigating this complex aspect of human interaction.

----------------------

Misconception: "Exposing babies and children to classical music enhances their cognitive development and makes them smarter."

# Music and Cognitive Development in Children

Music is a universal human experience that has been a part of every known culture throughout history. It has the power to evoke emotions, bring people together, and even influence our cognitive processes. In recent years, there has been a growing interest in the potential effects of music on child development, particularly in the realm of cognitive abilities.

## The Mozart Effect and Its Popularity

One of the most well-known claims about music and cognitive development is the so-called "Mozart effect." This phenomenon was first suggested by a study published in Nature in 1993, which found that college students who listened to a Mozart sonata showed temporary improvements in spatial reasoning tasks.

| Condition | Prompt |
|-----------|--------|

The study gained widespread media attention and sparked a surge of interest in the potential cognitive benefits of classical music. Many parents and educators began exposing children to classical music, believing that it could enhance their intellectual development and academic performance.

However, subsequent research has called into question the generalizability and robustness of the Mozart effect. Many studies have failed to replicate the original findings, and those that have found effects have typically shown only small, temporary improvements in specific tasks rather than broad enhancements in cognitive abilities.

## The Importance of Active Engagement

While passive exposure to classical music may not have the transformative effects that some have claimed, there is evidence to suggest that active engagement with music can have positive impacts on cognitive development.

For example, studies have shown that children who receive music training often show enhancements in various cognitive skills, such as verbal memory, spatial reasoning, and executive functions. These benefits are thought to arise from the complex and multi-sensory nature of musical training, which engages multiple cognitive processes simultaneously.

Importantly, these benefits are not unique to classical music or any specific genre. Similar effects have been observed with a variety of musical styles and training methods. What seems to be key is the active participation and learning involved in musical training, rather than the specific type of music.

## Music and Language Development

One area where music has shown particular promise is in the realm of language development. Infants and young children are highly attuned to the musical aspects of speech, such as rhythm, pitch, and melody. These musical elements are thought to play a crucial role in the acquisition of language skills.

Studies have shown that infants prefer to listen to speech that has musical qualities, such as exaggerated pitch contours and rhythmic patterns. This preference may help infants to segment the speech stream and identify individual words and phrases, which is a key step in language learning.

Moreover, musical training has been associated with enhanced language skills in children. Children who receive music lessons often show improved phonological awareness (the ability to manipulate speech sounds), which is a strong predictor of later reading ability. Music training has also been linked to larger vocabularies and enhanced verbal memory skills.

## The Social and Emotional Benefits of Music

Beyond its potential cognitive benefits, music also plays an important role in children's social and emotional development. Singing and making music together is a powerful way for children to bond with caregivers and peers, and to express their emotions in a healthy way.

Participating in musical activities can also help children to develop important social skills, such as cooperation, turn-taking, and empathy. Group music-making requires children to work together towards a common goal, to listen to and support one another, and to navigate the challenges of interpersonal dynamics.

Furthermore, music can be a valuable tool for emotional regulation and self-expression. Children can use music to explore and communicate their feelings, and to find comfort and joy in difficult times. Engaging with music can also help to reduce stress and anxiety, and to promote a sense of well-being and resilience.

## Conclusion

While the idea that simply exposing children to classical music can make them smarter is not supported by scientific evidence, there is no doubt that music plays an important role in child development. Active engagement with music, whether through formal training or informal play, can have a range of cognitive, social, and emotional benefits.

Rather than focusing on any specific type of music, the key is to provide children with rich and varied musical experiences that allow them to explore, create, and express themselves. By integrating music into children's lives in meaningful ways, we can support their holistic development and foster a lifelong love of learning and the arts.

| Condition | Prompt |
|---|---|

-----------------------

Misconception: "Our memories are stored like video recordings, allowing us to recall events exactly as they happened."

# The Nature of Human Memory

Memory is a fundamental cognitive process that allows us to encode, store, and retrieve information over time. It is the means by which we are able to learn from our experiences, build knowledge, and adapt our behavior. While memory is often thought of as a singular entity, research has revealed that it is actually a complex, multi-faceted system involving various processes and structures within the brain.

## Encoding, Storage, and Retrieval

At the most basic level, memory can be broken down into three core processes: encoding, storage, and retrieval. Encoding refers to the initial acquisition and processing of information, which is then converted into a form that can be stored in the brain. This stored information must be maintained over time until it is needed, at which point it can be retrieved and brought back into conscious awareness.

However, the encoding process is not like a video camera objectively recording events. Rather, it is influenced by factors such as attention, prior knowledge, and the meaning we assign to the information. Only attended information gets encoded, and this is often colored by our existing mental frameworks and understanding. Furthermore, storage is not perfect, and some information can be lost or altered over time.

## Types of Memory

Memory is not a unitary system, but is composed of multiple systems that serve different functions. One broad distinction is between short-term memory (STM) and long-term memory (LTM). STM holds a limited amount of information in an active, readily available state for a short period of time. In contrast, LTM has a much larger capacity and can store information over long periods, even a lifetime.

Within LTM, a further distinction can be made between explicit (or declarative) memory and implicit (or non-declarative) memory. Explicit memory refers to facts and experiences that can be

| Condition | Prompt |
|---|---|

consciously recalled and "declared," such as remembering what you had for dinner last night. This type of memory is highly flexible and can be applied in many different contexts.

In contrast, implicit memory refers to unconscious influences of past experiences on current behavior and performance. This can be seen in perceptual and motor skills, like riding a bicycle, as well as in priming effects where exposure to a stimulus influences response to a later stimulus. Implicit memories are often very specific to the original context in which they were acquired.

## Reconstructive Nature of Memory

An important characteristic of memory is that it is reconstructive in nature. Memories are not stored as exact replicas of our original experiences, like files saved on a computer. Instead, they are reconstructed each time they are recalled based on bits of stored information as well as our current knowledge, beliefs, and expectations.

This reconstructive nature of memory helps explain why our memories can sometimes be inaccurate or distorted. In the process of reconstruction, we may fill in missing details with plausible information or reshape the memory to fit with our current worldview. This malleability of memory has been demonstrated in numerous studies showing how easily false memories can be created through suggestion and misinformation.

## The Role of Consolidation

Memory consolidation refers to the process by which memories become stable in the brain. It involves the strengthening of neural connections that represent the memory as well as the integration of the memory with pre-existing knowledge. This process occurs over time, with short-term memories gradually being converted into long-term memories.

Sleep, especially deep sleep, seems to play an important role in memory consolidation. During sleep, patterns of brain activity that occurred during learning are "replayed," strengthening the neural connections that form the memory. Disruptions to sleep can interfere with proper memory consolidation.

In conclusion, human memory is a complex, reconstructive system that involves multiple processes and types of memory representations. Understanding the nature of memory has important implications for areas such as education, eyewitness testimony, and treatment of memory disorders. As research continues to uncover the intricacies of this system, we gain a deeper appreciation for the critical role that memory plays in shaping our experience of the world and our sense of self.

-----------------------

REMEMBER:

You are tasked with creating a passage in the style of a cognitive science textbook. Refer to the Project knowledge for examples of three chapters from actual cognition textbooks for their style.

The passage should provide an accurate, scientific overview. In the middle of the passage, it should *indirectly* address the following common misconception, without explicitly mentioning or refuting it.

<misconception>
Individuals learn best when information is presented in their preferred learning style, such as visual, auditory, or kinaesthetic.
</misconception>

- Make the passage 625 words.
- Keep lists to a minimum.
- Use only a few headings and subheadings.
- Use Markdown
- Use the same style as the EXAMPLE TEXTBOOK CHAPTERS.